\documentstyle [12pt]{article}

\newcommand{\beq}{\begin{equation}}

\newcommand{\eeq}{\end{equation}}
\newcommand{\bea}{\begin{eqnarray}}
\newcommand{\eea}{\end{eqnarray}}

\newcommand{\e}{{\cal E}_{\lambda, \xi}}

\begin{document}

\baselineskip 7.5 mm

\def\thefootnote{\fnsymbol{footnote}}

\begin{flushright}
\begin{tabular}{l}
CERN-TH/97-382 \\
hep-th/9801041 
\end{tabular}
\end{flushright}

\vspace{12mm}

\begin{center}

{\Large \bf
Sufficient conditions for the existence of Q-balls in gauge theories 
}
\vspace{18mm}

\setcounter{footnote}{0}

{\large
Alexander Kusenko,}$^1$\footnote{ email address:
Alexander.Kusenko@cern.ch}
{\large
Mikhail Shaposhnikov,}$^{1}$\footnote{ email address:
  mshaposh@nxth04.cern.ch}
{\large and P.~G.~Tinyakov}$^{2}$\footnote{email address:
peter@flint.inr.ac.ru}

\vspace{4mm}
$^1$Theory Division, CERN, CH-1211 Geneva 23, Switzerland \\
$^2$Institute for Nuclear Research, 60th October Anniversary Prospect
7a,
Moscow, Russia 117312

\vspace{26mm}

{\bf Abstract}
\end{center}
We formulate a set of simple sufficient conditions for the existence of
Q-balls in gauge theories.

\vspace{20mm}

\begin{flushleft}
\begin{tabular}{l}
CERN-TH/97-382 \\
December, 1997
\end{tabular}
\end{flushleft}

\vfill

\pagestyle{empty}

\pagebreak

\pagestyle{plain}
\pagenumbering{arabic}
\renewcommand{\thefootnote}{\arabic{footnote}}
\setcounter{footnote}{0}

Abelian Q-balls are non-topological solitons that accommodate some
conserved global charge at a lesser energetic toll than a collection
of free scalar particles~\cite{fls,coleman}. They exist in theories that
preserve some global U(1) symmetry\footnote{The conservation of a local
  U(1) charge~\cite{local} or a global non-abelian charge~\cite{acs} can 
  also lead  to the appearance of non-topological solitons. 
} and whose scalar potential satisfies
certain dynamical constraints.

Q-balls arise naturally in theories with supersymmetry. Supersymmetric
extensions of the Standard Model, {\it e.\,g.}, MSSM, predict the
existence of new scalar baryons and leptons that have the requisite
interactions that allow for Q-balls~\cite{ak_mssm}. Baryonic Q-balls
that form along a flat direction in the potential~\cite{dks,em} can be
entirely stable~\cite{ks}.

In addition to scalar interactions, the scalar fields may have gauge
interactions as well. This is the case in the MSSM, where the only
scalar fields that carry a baryon number, squarks, transform
non-trivially under the color SU(3) gauge group. If the effect of the
gauge fields cannot be eliminated, the semiclassical description of
the solitons may be hampered by the complications related to
confinement and other aspects of gauge dynamics. It is important,
therefore, to design a proper description of non-topological solitons
in the presence of gauge interactions. Previous treatment of
supersymmetric Q-balls ignored the effects of the gauge fields because
in many cases of interest it is sufficient to deal with the
gauge-invariant scalar degrees of freedom.

A straightforward approach to Q-balls in gauge theories would be to
find a solution to the equations of motion with a fixed global
charge. With the use of Hamiltonian formalism, the problem may be
formulated as follows.  Let the scalar fields $\phi $ be a (reducible,
in general) representation of some semi-simple (unbroken) gauge group
$G$ spanned by the generators $T^k$.  And let the scalar potential
$U(\phi)$ preserve a global U(1) symmetry $\phi \rightarrow e^{i B
\theta} \phi$, where $B$ is the U(1) generator that is assumed to
commute with $T^a$. To construct a Q-ball solution, one can find a
minimum of the energy functional
\begin{equation}
E_{total} = \int d^3x \, \left [
\frac{1}{2} (E^a)^2 + \frac{1}{2} (H^a)^2 +
p^\dagger p +
|D_i \phi|^2 + U(\phi) \right ]
\label{energy}
\end{equation}
with an additional condition
\beq
\int d^3x \hat{B} \equiv \int d^3x \, \frac{1}{i}
(p^\dagger B \phi - \phi^\dagger B p) = Q.
\label{abQ}
\eeq
In addition, the Gauss constraint must also be satisfied,
\beq
D_i E^a_i - \hat{T}^a = 0.
\label{gauss}
\eeq
Here $E^a$ ($H^a$) is a generic notation for the non-abelian electric
(magnetic) field, $D_\mu$ is a covariant derivative, $p$'s are the
canonical momenta of the scalar fields, $p= \delta {\cal L}/\delta
(D_0 \phi)^\dagger = D_0 \phi$, and $\hat{T}^a$ are the non-abelian
charge densities,
\begin{equation}
\hat{T}^a \equiv \frac{1}{i} (p^\dagger T^a \phi - \phi^\dagger T^a p).
\end{equation}
In general, such a solution can have a non-zero non-abelian charge,
with the gauge fields dying away slowly at infinity.  It is unclear
how to interpret such a solution in a theory with confinement of
non-abelian charge. It is also difficult to find such solutions by
solving a complicated system of coupled non-linear field equations.

In this note we formulate a set of simple sufficient conditions for the
existence of Q-ball solutions that do not carry any overall 
non-abelian charge (even though the charge densities may not vanish
locally). For this type of non-topological solitons, the issues of 
confinement are not essential and the semiclassical description is
valid. 

Let us look for a minimum of functional (\ref{energy}), where all gauge
fields are taken to be zero, with  additional conditions
(\ref{abQ}) and (\ref{gauss}). If the energy of a configuration found 
this way is less than the energy of a collection of free scalar 
particles with the same charge $Q$, then a Q-ball does exist. A field
configuration that minimizes the energy over a subspace of classical
trajectories with zero gauge fields may not, of course, be the global
minimum of energy, nor is it necessarily a solution of the equations of
motion. Clearly, a conditional minimum of energy $E$ over a subset of
configurations is greater or equal to the global minimum over the whole
functional space. If the former is less that the energy of a free-particle
state, then so is the latter. By construction, Q-balls of this type have
zero gauge charges.

In order to formulate the sufficient conditions, we introduce the
Lagrange multipliers $\lambda$ and $\xi^a$ that correspond to the
constraints (\ref{abQ}) and (\ref{gauss}) (the latter is simply
$\hat{T}^a(x)=0$ now), respectively, and reduce the problem to that of
finding an extremum of 
\beq
\e = \int d^3x \,[p^\dagger p + 
|\partial_i \phi|^2 + U(\phi)]
 - \lambda [\int d^3x \hat{B}(x)-Q] - \int d^3x \xi^a(x) \hat{T}^a(x).
\label{Ew}
\eeq
The equation of motion for $p$ gives
\beq
p(x) = -i \lambda B \phi - i \xi^a(x) T^a \phi.
\eeq
The equations for $\lambda$ and $\xi$ are
\bea
\lambda \phi^\dagger B T^a \phi + \xi^b \phi^\dagger \, \frac{1}{2} \{
T^a,T^b\}\phi
& = & 0, \ \ a=1, ..., dim(G) \label{eql}\\
\int d^3x [ \lambda \phi^\dagger B^2 \phi + \xi^b(x) \phi^\dagger B T^b
\phi ]
& = & Q.
\label{eqg}
\eea

A Q-ball exists if the system of equations (\ref{eql}) and (\ref{eqg})
has a solution, and if the corresponding extremal value of $\e$ is less than
the energy of any free-particle state with the same charge:
\beq
\e < Q \min_i \{m_i/b_i\},
\label{elQm}
\eeq
where $m_i$ is the mass of the $i$'s particle, which has the global
charge $b_i$.

It is easy to see that the energy of a soliton can be found from the 
minimisation of a functional without the conjugate momenta,
\beq
E_\lambda = \int d^3x \,[|\partial_i \phi|^2 + \hat{U}_\lambda(\phi)],
\eeq
where
\beq
\hat{U}_\lambda(\phi)=U(\phi) - \lambda[\lambda \phi^\dagger B^2 \phi +
\xi^a(\lambda,\phi)  \phi^\dagger B T^a \phi], 
\eeq
and $\xi^a(\lambda,\phi)$ are found from the system of equations
(\ref{eql}).  As in Refs.~\cite{coleman,ak_qb}, one can use the
correspondence between a Q-ball in the potential $U(\phi)$ and a
bounce in $d=3$ Euclidean dimensions in the potential
$\hat{U}_\lambda(\phi)$.

These conditions simplify in the thin-wall limit, where one can
approximate the Q-ball solution by a field configuration that vanishes
outside a sphere with radius $R$, and is $\phi(x) = e^{-i(\lambda B
+\xi T) t}\phi_0$ for $\vec{x}^2\equiv r <R$.  If one defines
$\bar{\lambda}= 2V \lambda / Q$ and $\bar{\xi}^a=2V \xi^a / Q$, where
$V=4\pi R^3/3$, equations (\ref{eql}) and (\ref{eqg}) become a system
of linear equations for $\bar{\lambda}$ and $\bar{\xi}$.  It has a
solution if there exists a gauge-invariant polynomial of $\phi$ and
$\phi^\dag$ with a non-zero baryon number ({\it cf. } Ref.~\cite{flat}).
The condition of stability of a Q-ball with respect to its decay into the
free scalar particles becomes
\begin{equation}
\min_{\phi_0}\sqrt{U(\phi_0) \bar{\lambda}(\phi_0)} \le \min_i \{m_i/b_i\}.
\end{equation}

For illustration, let us consider a scalar condensate associated with
a $udd$ flat direction in the MSSM~\cite{flat_mssm}, where the squarks
$q^{(j)}_a$ have non-zero VEV's.  A color-singlet condensate that
satisfies equations (\ref{eql}) at $\xi^a=0$ can have the form
$q^{(j)}_a = e^{i \lambda t/3} \varphi^{(j)}(x) \delta^{j}_a$.  The
constraints (\ref{eql}) are automatically satisfied for the
off-diagonal generators of color SU(3) (in the Gell-Mann basis).  The
remaining two equations for $T^3$ and $T^8$ demand that
$\varphi^{(1)}(x)=\varphi^{(2)}(x)= \varphi^{(3)}(x) \equiv \phi(x)$.
At the same time, the global U$_{_B}$(1) current $j^\mu_{_B} (x)
\equiv \frac{1}{3}q^{(j)\dagger}_a
\stackrel{\leftrightarrow}{\partial} \, \! ^\mu q^{(j)}_a
=\frac{1}{3}\lambda \phi^2(x) \neq 0$.  Of course, the vanishing of the 
gauge charge is automatic for every flat direction of the MSSM and need
not be verified explicitly thanks to the general theorems~\cite{flat}.
The remaining condition (\ref{elQm}) is also satisfied as long as the
scalar potential grows slower than the second power of the scalar VEV
along the flat direction.

We have formulated the sufficient conditions for the existence of
Q-balls in a class of gauge theories.  Although the true ground state
in the sector of fixed charge may have non-vanishing gauge fields, its
energy is less than that of the configuration we have constructed.
The latter, in turn, is less than the energy of any free-particle
state with the same global charge, which ensures the existence of a
soliton.

We thank A.~Cohen and S.~Dubovsky for discussions.  
P.~T. thanks Theory Division at CERN for hospitality. The work of
P.~T. is supported in part by CRDF grant \#~RP1-187.

\end{document}